\documentclass[conference]{IEEEtran}

\usepackage{cite}
\usepackage{amsmath,amssymb,amsfonts}
\usepackage{algorithmic}
\usepackage{graphicx}
\usepackage{textcomp}
\usepackage{xcolor}
\def\BibTeX{{\rm B\kern-.05em{\sc i\kern-.025em b}\kern-.08em
    T\kern-.1667em\lower.7ex\hbox{E}\kern-.125emX}}
\usepackage[linesnumbered,ruled,vlined]{algorithm2e}
\usepackage{multirow}
\usepackage{xspace}
\usepackage{enumitem}
\usepackage{tikz}
\usepackage{cleveref} 
\usepackage{adjustbox}
\usepackage{subcaption}
\usepackage{tcolorbox}
\usepackage{booktabs}

\begin{document}

\title{Towards More Trustworthy Deep Code Models by Enabling Out-of-Distribution Detection}

\author{\IEEEauthorblockN{%
    Yanfu Yan\textsuperscript{\textsection}, 
    Viet Duong\textsuperscript{\textsection}, 
    Huajie Shao, 
    Denys Poshyvanyk}
\IEEEauthorblockA{Department of Computer Science,
William \& Mary\\
Williamsburg, Virginia, USA\\
Email: yyan09, vqduong, hshao, dposhyvanyk \{@wm.edu\}}}

\pagestyle{plain}

\input{macros}
\maketitle
\begingroup\renewcommand\thefootnote{\textsection}
\footnotetext{Equal contribution.}
\endgroup

\begin{abstract}

Numerous machine learning (ML) models have been developed, including those for software engineering (SE) tasks, under the assumption that training and testing data come from the same distribution. However, training and testing distributions often differ, as training datasets rarely encompass the entire distribution, while testing distribution tends to shift over time. Hence, when confronted with out-of-distribution (OOD) instances that differ from the training data, a reliable and trustworthy SE ML model must be capable of detecting them to either abstain from making predictions, or potentially forward these OODs to appropriate models handling other categories or tasks. 

In this paper, we develop two types of SE-specific OOD detection models, unsupervised and weakly-supervised OOD detection for code. The unsupervised OOD detection approach is trained solely on in-distribution samples while the weakly-supervised approach utilizes a tiny number of OOD samples to further enhance the detection performance in various OOD scenarios. Extensive experimental results demonstrate that our proposed methods significantly outperform the baselines in detecting OOD samples from four different scenarios simultaneously and also positively impact a main code understanding task.

\end{abstract}

\section{Introduction}\label{sec:intro}
Extensive ML models have been developed under the assumption that training and testing data come from the same distribution (\ie \textit{closed-world assumption}). However, this assumption is often violated in practice, where deployed models may frequently encounter out-of-distribution (OOD) instances that are not seen in training ~\cite{torralba2011unbiased}. 
For instance, a model trained on high-quality code may struggle to comprehend buggy code. Adapting ML models to distribution shifts is possible but challenging and costly due to the constantly evolving data~\cite{liu2022deep}. Moreover, even if the training data is up-to-date, models will still encounter unforeseen scenarios under the open-world setting. Failure to recognize an OOD sample, and consequently to produce incorrect predictions, significantly compromises the reliability of a model. A reliable and trustworthy ML model should not only achieve high performance on samples from known distributions, \ie in-distribution (ID) data, but also accurately detect OOD samples which can then either abstain from making predictions, or potentially be forwarded to appropriate models handling other distributions or tasks.

OOD detection has been extensively studied in computer vision (CV)~\cite{yang2021generalized} and natural language processing (NLP)~\cite{li2023survey} across a range of tasks (\eg image/sentiment classification, question answering). Existing OOD detectors typically design a scoring function to derive confidence/ID scores, enabling the detection of OOD samples based on a predefined threshold. These OOD detectors serve as an auxiliary function to the original ML models and ensures a high proportion (\eg 95\%)~\cite{ming2022delving} of ID data finally retained based on the threshold. This is crucial to prevent the OOD auxiliary scoring from adversely affecting ML models' performance on their main image/language-related tasks. Current OOD detection approaches are proposed in supervised, unsupervised, and weakly-supervised regimes depending on the availability of OOD data. Supervised approaches~\cite{hendrycks2018deep} learn a classical binary classifier based on both ID and OOD data, but in practice, it is hard to assume the presence of a large dataset that captures everything different from the ID data. Unsupervised ones~\cite{mai2022self, zhou2021contrastive} only utilize ID data for training, but are likely to suffer from poor performance. Recent studies have demonstrated that weak supervision~\cite{tian2020few, majhi2021weakly, kim2023key, kim-etal-2023-pseudo} can remarkably outperform unsupervised learning methods for anomaly/OOD detection. Some weakly-supervised approaches~\cite{majhi2021weakly, kim2023key, kim-etal-2023-pseudo} generate pseudo-labeled OODs by partially corrupting ID data based on output attention mappings, while others~\cite{tian2020few} leverage a tiny collection of labeled OODs (\eg 1\% of ID data) to detect specific OOD types in the applications where access to OOD samples is limited and pseudo OOD generation is challenging~\cite{yoo2022data}. However, none of these ML approaches have been applied in the context of SE for code-related tasks.

Existing OOD detection research in SE primarily focuses on anomaly detection or software defect detection. Anomaly detection techniques~\cite{le2022log, wang2023logonline, traces2020guo, traces2021liu} are designed to detect anomalous system states (\eg failed processes, availability issues, security incidents) during system running based on \textit{monitoring data} (\eg logs, traces), but they still cannot been applied to the code context. There also exists a body of research dedicated to detecting suspicious defects in \textit{source code} (\eg vulnerability detection~\cite{sejfia2024toward, steenhoek2024dataflow, vul2024cao}, neural bug detection~\cite{ allamanis2021self, he2022distribution}). Although defective source code represents a type of distribution shifts from normal code, current defect detection techniques are not sufficient to cover a broad range of unseen scenarios considered by OOD detection.

Therefore, the goal of this work is to address the OOD detection problem in the context of SE for code-related tasks. While Transformer-based~\cite{vaswani2017attention} NL-PL (programming language) models have shown remarkable success in code understanding and generation ~\cite{guo2020graphcodebert, guo2022unixcoder, athena} by utilizing bimodal data (\ie comment and code), they often assume training and testing examples belong to the same distribution. Thus, these models may not guarantee the robustness against OOD instances in the open world (as evidenced by~\cite{hendrycks2020pretrained} for NL Transformers). For instance, a code search engine, which is trained on GitHub comment-based queries and code, is likely to fail in user questions and code answers from StackOverflow. 

In this paper, we systematically investigate the ability of pre-trained NL-PL models~\cite{guo2020graphcodebert, guo2022unixcoder, liu2023contrabert} in detecting OOD instances and the impact of OOD detection on a downstream code task (\ie code search). While NLP OOD detection techniques show promise for adaptation to NL-PL models due to the similarity between NL and PL, they can only detect textual OODs from uni-modal data. However, in the SE context for code-related tasks, distribution shifts can occur in either modality (comment or code) or both of them. An effective OOD code detector should be able to detect OOD from comments, code, or both modalities, by utilizing multi-modal NL-PL pairs. Several multi-modal approaches have been proposed for vision OOD detection~\cite{ming2022delving, esmaeilpour2022zero}, utilizing information from both images and their textual descriptions, but they are still designed to detect \textit{only} visual OODs.





To overcome these challenges, we develop two types of multi-modal OOD detection models to equip NL-PL models with OOD code detection capability. The first one is unsupervised (coined as COOD), which fine-tunes the NL-PL models to closely align NL-PL representations solely from ID data~\cite{husain2019codesearchnet} based on the multi-modal contrastive learning~\cite{oord2018representation}, and then uses their prediction confidences as OOD scores. The contrastive learning objective is expected to effectively capture high-level alignment information within (NL, PL) pairs to detect OODs. To further enhance the OOD detection performance, we propose a weakly-supervised OOD detection model, COOD+, which utilizes a tiny collection of OOD samples (\eg 1\%) during model training. Current techniques in ML typically considered unsupervised contrastive learning~\cite{zhou2021contrastive} or outlier exposure~\cite{hendrycks2018deep, liu2020energy}, in conjunction with a scoring function, limiting their ability to detect OODs from just one modality. In contrast, our COOD+ integrates an improved contrastive learning module with a binary OOD rejection module in order to effectively detect OODs from NL, PL, or both modalities. OOD samples are then identified by a combination of two different scoring functions: the confidence scores produced by the contrastive learning module and the prediction probabilities of the binary OOD rejection module.

Due to the lack of evaluation benchmarks for OOD code detection, we create a new benchmark tailored for code context following the construction principles in ML~\cite{zhou2021contrastive, mai2022self}, but containing more OOD scenarios: (1) aligned (NL, PL) pairs collected from a new domain, \eg from StackOverflow rather than GitHub, (2) misaligned (NL, PL) pairs,  (3) the presence of syntactic errors in NL descriptions, and (4) buggy source code. We first evaluate the proposed models on two real-world datasets, CodeSearchNet-Java and CodeSearchNet-Python, and establish a range of unsupervised and weakly-supervised baselines for comparison. Experimental results show that both COOD and COOD+ models significantly outperform the best unsupervised and weakly-supervised baselines, respectively. Specifically, our unsupervised COOD is moderately capable of detecting OODs from three scenarios but does not perform well across all four scenarios. By integrating two modules, our COOD+ model effectively detects OODs from all scenarios simultaneously.

Furthermore, we apply our approaches to improve the robustness of existing (NL, PL) models for the code search task under the four OOD scenarios described above. By corrupting 15\% of the testing dataset with OOD examples, we demonstrate that NL-PL models actually are not robust to OOD samples. Specifically, the performance of a fine-tuned GraphCodeBERT code search model drops by around 5\% due to the presence of OODs. Subsequently, we filter the corrupted testing dataset with our COOD/COOD+, and show that our detectors successfully recover this performance loss and also improve the code search performance compared to the original testing set. In summary, the contributions of this paper are:

\begin{itemize}[noitemsep, leftmargin=1.2em]  
\item A novel OOD benchmark specifically designed for code contexts, encompassing multiple OOD scenarios;

\item The first work to address OOD detection for code across four distinct scenarios;  

\item A multi-modal OOD detection framework for NL-PL pre-trained models, leveraging contrastive learning in both unsupervised and weakly-supervised settings;

\item A comprehensive evaluation showcasing the superior performance of our COOD and COOD+ frameworks in detecting OOD samples across four scenarios;

\item An online appendix providing the full codebase and experimental infrastructure of our approaches~\cite{cood-tool}. 

\end{itemize}
\section{Related Work}\label{sec:related}
We review the related work on OOD detection in various fields such as computer vision (CV), natural language processing (NLP), and software engineering (SE), and then point out unique characteristics of our approach. 

\subsection{OOD Detection in SE}

To ensure the reliability and safety of large-scale software systems, extensive work~\cite{ le2022log, yu2024deep, zhang2024metalog} has been conducted on anomaly detection to identify anomalous system state (\eg failed processes, availability issues, security incidents) during system running based on \texttt{\small monitoring data} (not in code format). Specifically, monitoring data includes logs~\cite{le2022log, wang2023logonline}, metrics (\eg response time, CPU usage)~\cite{metrics2021zervas}, traces~\cite{traces2020guo, traces2021liu}, etc. While some approaches utilize supervised learning techniques~\cite{zhang2019robust, lu2018detecting}, others employ unsupervised~\cite{farzad2020unsupervised} or semi-supervised learning~\cite{yang2021semi, Heter2023lee} due to insufficient anomaly labels. However, none of these anomaly detection techniques target code-based OOD detection, the main focus of our work. We mention this research line here since some existing OOD-related work in ML use the terms \textit{anomaly detection} and \textit{(generalized) OOD detection}  interchangeably~\cite{hendrycks2018deep}, but anomaly detection in SE has distinct characteristics as described above.


Additionally, current defect detection techniques~\cite{lu2021codexglue} in SE typically identify defects by analyzing \texttt{\small source code} with code semantic features extracted. Research in vulnerability detection focuses on security-related defects, such as buffer overflows and use-after-free. Compared to conventional static tools~\cite{CodeQLtool, Checkmarxtool}, DL-based techniques~\cite{sejfia2024toward, steenhoek2024dataflow, cul2024liu, vul2024cao} utilize Graph Neural Networks (GNNs) or Transformers to learn implicit vulnerability patterns from source code. Additionally, bug detection techniques~\cite{kanade2020learning, chen2021plur, allamanis2021self, he2022distribution} also fall under the umbrella of defect detection but typically address semantically-incorrect code (\eg wrong binary operators, variable misuse) which is not necessarily security-related and probably syntactically feasible. Although our focus is also on source code, defective code is only considered as one scenario within the scope of our OOD detection problem. 

More recently, several research has explored the robustness and generalization of source code models to different OOD scenarios~\cite{hu2023codes, weyssow2023usage, hajipour2024simscood}. Hu \etal introduced a benchmark dataset to assess the performance of code models under distribution shifts~\cite{hu2023codes}, while others investigated fine-tuning strategies like low-rank adaptation~\cite{hajipour2024simscood} and continual learning~\cite{weyssow2023usage} for enhanced \textit{generalization} on OOD data. However, these studies did not specifically tackle OOD \textit{detection}, and existing unsupervised OOD detectors have shown limited effectiveness for source code data~\cite{hu2023codes}. In short, unlike prior work, our study directly aims to improve the OOD detection performance of existing code-related models, ensuring greater robustness and trustworthiness in the open world where many unseen OOD scenarios may be encountered.

\subsection{OOD Detection in CV and NLP}

In the ML community, OOD detection~\cite{hendrycksbaseline, yang2021generalized, salehiunified, li2023survey} has been extensively studied over the years, leading to a better-defined and formulated task. The primary objective of OOD detection here is to design an auxiliary ID-OOD classifier derived from neural-based visual and/or textual models based on OOD scores. Given that correctly predicted instances tend to have greater maximum softmax probabilities (MSP) than incorrectly predicted and OOD instances, MSP-based OOD scoring function~\cite{hendrycksbaseline, hein2019relu} weree initially utilized to identify OOD samples. Subsequently, energy- and distance-based scores~\cite{liu2020energy, zhou2021contrastive,sun2022knn} have also been utilized to derive OOD scores. For visual OOD data, existing techniques often aim for multi-class classification tasks (\eg image classification) and learn a $K+1$ classifier assuming that the unseen space is included in the additional class~\cite{hu2021outlier, liznerski2022exposing}. The OOD data utilized for evaluation is typically constructed from a completely different dataset (out-domain data) or by holding out a subset of classes in a categorized dataset, where one category is considered normal and the remaining categories are treated as OOD.

In the context of textual data, OOD detection techniques are applied to both classification tasks (\eg sentiment/topic classification~\cite{zhou2021contrastive, kim-etal-2023-pseudo}) and selective prediction tasks~\cite{kamath-etal-2020-selective,varshney-etal-2022-investigating, xin2021art} (\eg question answering, semantic equivalence judgments). These techniques rely on various algorithmic solutions including outlier exposure~\cite{zeng2021adversarial, hu2021outlier}, data augmentation~\cite{zheng2020aug, zhan-etal-2021-scope}, contrastive learning~\cite{zhou2021contrastive, 10.1109/TASLP.2022.3162081}, \etc. Compared to traditional neural-based language models, pre-trained Transformer-based~\cite{vaswani2017attention} models exhibit greater robustness to distributional shifts and are more effective in identifying OOD instances~\cite{hendrycks2020pretrained, xu-etal-2021-unsupervised}. Besides the out-domain data, text-based OOD detection also consider syntactic OOD data~\cite{mai2022self} due to the intrinsic characteristics of sentences. Syntactic OOD and ID data come from the same domain, but the syntactic OOD data has its word order shuffled, which allows for the measurement of OOD detectors' sensitivity to underlying syntactic information while preserving word frequency.

Some studies~\cite{sun2020real, wang2021radar} have explored the incorporation of multi-modal data into neural-based models to improve OOD detection accuracy. Recently, CLIP-based methods~\cite{fort2021exploring, ming2022delving,esmaeilpour2022zero} have emerged as a promising approach for OOD detection by leveraging vision-language bimodal data, exhibiting superior performance over uni-modal data only. The main intuition behind these approaches is to take advantage of the alignment between visual classes or concepts and their textual descriptions. For instance, Ming et al.~\cite{ming2022delving} detect visual OOD in an unsupervised manner by matching visual features with known ID concepts in the corresponding textual descriptions. 

However, these studies typically focus on detecting OOD data from at most two scenarios (\ie out-domain and shuffled-text OODs) within a \textit{single} modality. Even multi-modal approaches are often limited to detecting only visual OODs by additionally considering accompanying textual descriptions. Our proposed approach aims to effectively identify OOD samples from four distinct scenarios across \textit{two} modalities (\ie NL and PL). To achieve this, we utilize a combination of different scoring functions from two different modules: cosine similarities of a contrastive learning module and prediction probabilities of a binary OOD classifier. 




\section{Preliminaries}
Below we review some background knowledge and techniques that will be used in the proposed approach.

\subsection{Code Representation Learning} 

Representation learning refers to the process of learning a parametric mapping from the raw input data domain to a low-dimensional latent space, aiming to extract more abstract and useful concepts that can improve performance on a range of downstream tasks. Recently, self-supervised representation learning has become prominent in the ML community due to the success of large pre-trained models. The similarity between NL and PL has led to the development of numerous NL-PL pre-trained models~\cite{guo2020graphcodebert, guo2022unixcoder}, resulting in significant improvements across various code-related tasks such as code search, generation, and clone detection. CodeBERT\cite{feng2020codebert} stands as the first Transformer-based NL-PL pre-trained model, while GraphCodeBERT~\cite{guo2020graphcodebert} further enhances its performance by considering the inherent structure of code (\ie data flow) during pre-training. Subsequent models like UniXcoder~\cite{guo2022unixcoder} and ContraBERT~\cite{liu2023contrabert} further enhance the code understanding capabilities. In our research study, we are concerned with extending self-supervised NL-PL representation learning to the OOD detection task, which is outside the scope of existing pre-trained models, but is crucial for ensuring the trustworthiness of these models in real-world applications. 

\subsection{Multimodal Contrastive Learning}

Contrastive learning~\cite{1640964} is an emerging technique that learns discriminative representations from data that are organized into similar/dissimilar pairs. It has been developed over multiple sub-fields, such as NLP, CV \cite{khosla2020supervised, chen2020simple}, and SE~\cite{jain2020contrastive, guo2022unixcoder}. Recently, researchers have developed multi-modal approaches that combine contrastive learning with multiple data modalities, achieving superior performance over uni-modal modals in various tasks~\cite{radford2021learning, liu2023contrabert}. Due to the prowess of contrastive learning in discriminative tasks (\eg classification and information retrieval), it naturally fits the OOD detection domain, as shown by studies on both uni-modal and multi-modal data~\cite{ming2022delving, qiu2022unsupervised, esmaeilpour2022zero}. Inspired by this, we apply contrastive learning to NL-PL data to learn representative and discriminate features for both ID and OOD instances and transfer this knowledge to train our OOD detector.

\section{Approach}\label{sec:approach} 

In this section, we first formally define the OOD code detection problem for (NL, PL) models (Sec. IV-A), then introduce the overall proposed framework (Sec. IV-B), and finally present details of unsupervised COOD and weakly-supervised COOD+ in Sec. IV-C and Sec. IV-D, respectively.
\subsection{Problem Statement}

Since current state-of-the-art code-related models~\cite{guo2020graphcodebert, guo2022unixcoder} typically extract code semantics by capturing the semantic connection between NL (\ie comment) and PL (\ie code) modalities, we formally defined OOD samples involving these two modalities in the SE context by following the convention in ML~\cite{yang2021generalized, zhou2021contrastive}. Consider a dataset comprising training samples $((t_1, c_1), y_1), ((t_2, c_2), y_2), ...$ from the joint distribution $P((T, C),Y)$ over the space $\mathcal{(T, C)}\times \mathcal{Y}$, and a neural-based code model is trained to learn this distribution. Here, $((t_1, c_1), y_1)$ represents the first input pair of (comment, code) along with its ground-truth prediction in the training corpus. $T$, $C$ and $Y$ are random variables on an input (comment, code) space $\mathcal{(T, C)}$ and a output (semantic) space $\mathcal{Y}$, respectively. OOD code samples refer to instances that typically deviate from the overall training distribution due to distribution shifts. The concept of distribution shift is very \textit{broad}~\cite{yang2021generalized, salehiunified} and can occur in either the marginal distribution $P(T, C)$, or both $P(Y)$ and $P(T, C)$.

We then formally define the OOD code detection task following~\cite{hsu2020generalized, liu2020energy, tian2020few, kim-etal-2023-pseudo} as follows. Given a main code-related task (\eg clone detection, code search, \etc), the objective here is to develop an \textit{auxiliary} scoring function $g: \mathcal{(T, C)} \rightarrow \mathcal{R}$ that assigns higher scores to normal instances where $ ((t, c), y) \in P((T, C),Y)$, and lower scores to OOD instances where $((t, c), y) \notin P((T, C),Y)$. Based on whether to use OOD instances during the main-task training of pre-trained NL-PL models, we define OOD for code in two settings, namely unsupervised and weakly-supervised learning. For the unsupervised setting, only normal data is used in the main-task training. Conversely, weakly-supervised approaches utilize ID and a tiny collection of OOD data (\eg 1\% of ID data)~\cite{tian2020few} in training. In this context, the output space $\mathcal{Y}$ is typically a binary set, indicating normal or abnormal, which is probably unknown during inference. Due to the small number of training OOD data, the OOD samples required by our COOD+ and other existing weakly-supervised approaches~\cite{ruff2020deep, tian2020few} in ML can be generated at minimal cost and feasibly verified by human experts when necessary.




\subsection{Overview}
\begin{figure*}[!thb]
\begin{center}
 \includegraphics[width=0.98\textwidth]{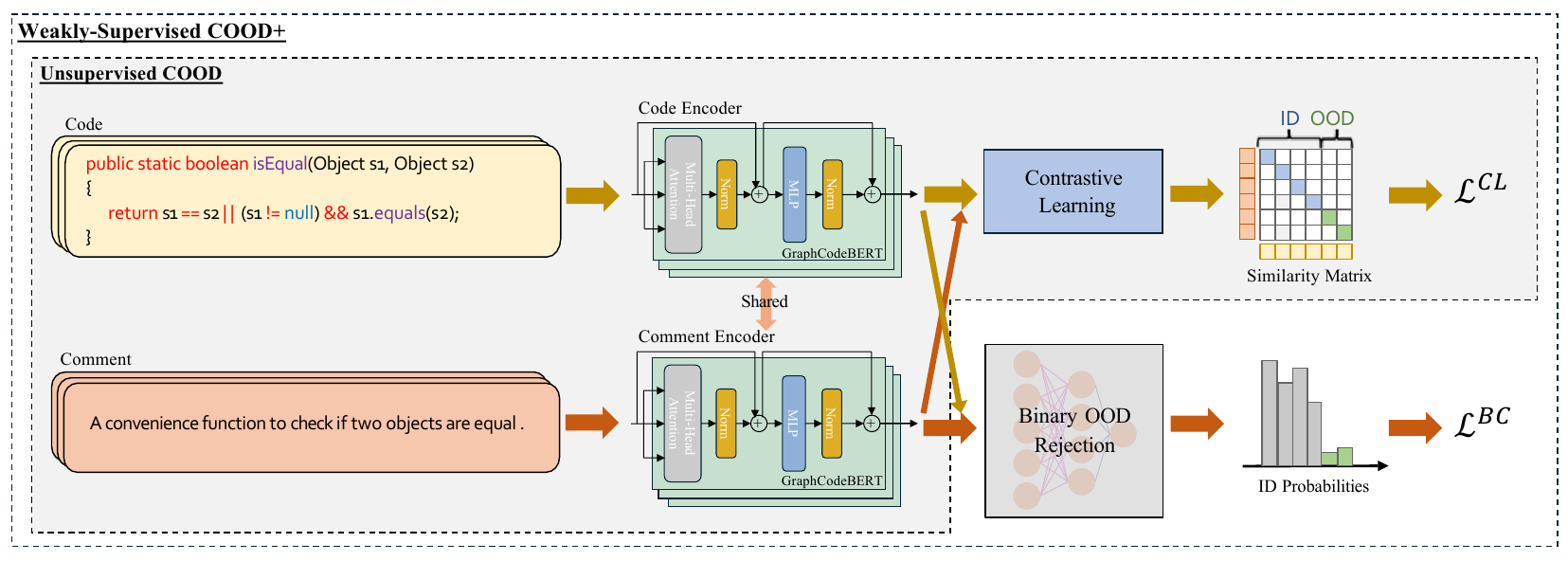}
    \caption{The Overview of Our Proposed COOD and COOD+ Approaches for OOD Detection 
    }
    \label{fig:cad}
\end{center}

\end{figure*}

Overall, there are two versions of our COOD approach: unsupervised COOD and weakly-supervised COOD+. Given a multi-modal (NL, PL) input, the unsupervised COOD learns distinct representations based on a contrastive learning module by utilizing a pre-trained Transformer-based code representation model (\ie GraphCodeBERT~\cite{guo2020graphcodebert}). Then, these representations are mapped to distance-based OOD detection scores in order to indicate whether the test samples are OODs during inference. The weakly-supervised COOD+ further integrates a improved contrastive learning module with a binary OOD rejection module to enhance the detection performance by using a very tiny number of OOD data during model training. The OOD samples are then identified by the detection scores produced by the contrastive learning module as well as the prediction probabilities of the binary OOD rejection module. 

\subsection{Unsupervised COOD}

Our unsupervised COOD approach consists of a contrastive learning (CL) module trained only on ID samples. Specifically, given (comment, code) pairs as input, we fine-tune a comment encoder and a code encoder through a contrastive objective to learn discriminative features, which are expected to help identify OOD samples based on a scoring function. 

The (comment, code) pairs are first converted into the comment and code representations, which are processed by the comment and code encoder, respectively. We use the pre-trained GraphCodeBERT model~\cite{guo2020graphcodebert} as the encoder architecture (\ie backbone). GraphCodeBERT is a Transformer-based model pre-trained on six PLs by taking the (comment, code) pairs as well as the data flow graph of the code as input, which has shown superior performance on code understanding and generation tasks. All the representations of the last hidden states of the GraphCodeBERT encoder are averaged to obtain the sequence-level features of comment and code.

\textbf{Contrastive Learning Module.} To achieve the contrastive learning objective, we fine-tune the base (GraphCodeBERT) encoders with the InfoNCE loss~\cite{oord2018representation}. The comment and code encoders follow the Siamese architecture~\cite{guo2022unixcoder} since they are designed to be identical subnetworks with the same GraphCodeBERT backbones, in which their parameters (\ie weights and biases) are shared during fine-tuning. Parameter sharing can reduce the model size and has shown state-of-the-art performance for the code search task~\cite{shi2023cocosoda}. To extract discriminative features for (comment, code) pairs, we organize them into functionally-similar positive pairs and dissimilar negative (unpaired) pairs. Through a contrastive objective, positive pairs are drawn together, while unpaired comment and code are pulled apart. Specifically, for each positive (comment, code) pair $(t_i, c_i)$ in the batch, the code in each of other pairs and $t_i$ are constructed as in-batch negatives, similarly for the comment side. The loss function then formulates the contrastive learning as a classification task, which maximizes the probability of selecting positives along the diagonal of the similarity matrix (as shown in Fig. \ref{fig:cad}) by taking the \textit{softmax} of projected embedding similarities across the batch.  The loss function can be summarized as follows:

\begin{align}\small
\label{eq:infonce}
\begin{split}
    \mathcal{L}^{CL} &= -\frac{1}{2N}(\sum_{n=1}^{N}\log \frac{e^{sim(v_{t_i},v_{c_i})/\tau}}{\sum_{j=1}^{N}e^{sim(v_{t_i},v_{c_j})/\tau}}\\ &+ \sum_{n=1}^{N}\log \frac{e^{sim(v_{t_i},v_{c_i})/\tau}}{\sum_{j=1}^{N}e^{sim(v_{t_j},v_{c_i})/\tau}})
\end{split}
\end{align}
where $v_{t_i}$ and $v_{c_i}$ represent the extracted features of the comment $t_i$ and the code ${c_i}$. $\tau$ is the temperature hyperparameter, which is set to 0.07 following previous work~\cite{shi2023cocosoda}.  $sim(v_{c_i},v_{t_i})$ and $sim(v_{t_i},v_{c_j})/sim(v_{t_j},v_{c_i})$ represent the cosine similarities between comment and code features for positive and negative pairs, respectively. $N$ is the number of input pairs in the batch. InfoNCE loss is designed for self-supervised learning and learns to distinguish positive pairs from in-batch negatives. Compared to other contrastive losses~\cite{khosla2020supervised, zhou2021contrastive}, it can take advantage of large batch size to automatically construct many diverse in-batch negatives for robustness representation learning, which is more effective to capture the alignment information between comment and code. 

\textbf{Scoring Function.} Existing OOD detection techniques in ML derive scoring functions based on model's output, which typically map the learned class-probabilistic distributions to OOD detection scores for testing samples. Maximum Softmax Probability (MSP)~\cite{hendrycks2016baseline} is commonly used for OOD scoring. This method uses the maximum classification probability $\max_{l\in L}\textit{softmax}(f(v_{t}, v_{c}))$, where $f(v_{t}, v_{c})$ is the output of the classification model, with low scores indicating low likelihoods of being OOD. However, NL-PL code search models typically utilize the similarity retrieval scores of NL-PL output representations to make predictions. Therefore, to enable simultaneous similarity and OOD inference, we alternatively extract cosine similarity scores of testing NL-PL pairs as OOD detection scores, denoted as $P^{CL}=sim(v_{c},v_{t})$. The underlying intuition behind this scoring metric is that OOD testing samples should receive low retrieval confidence from the model fine-tuned on ID data, which establishes a closer relationship between ID (comment, code) pairs. Hence, this scoring function also assigns higher scores to ID data and lower scores to OOD data similar to previous scoring methods.

\subsection{Weakly-Supervised COOD+}

To further enhance the performance of unsupervised COOD, we extend it to a weakly-supervised detection model, called COOD+, which takes advantage of a few OOD examples. Inspired by~\cite{duong2024general}, our COOD+ combines an improved contrastive learning (CL) and a binary OOD rejection classifier (BC). The improved CL module adopts a margin-based loss~\cite{xue2009svm} which enforces a margin of difference between the cosine similarities of aligned and unaligned (comment, code) pairs, and constrains the cosine similarities of OOD pairs below another margin. The BC module integrates features from both comments and code to calculate the probabilities of OOD pairs. The OOD scoring function is then designed by combining the cosine similarity scores from the CL module and the prediction probabilities from the BC module. Below, we detail each component of our weakly-supervised COOD+.

\textbf{Improved Contrastive Learning (CL) Module.} Given a batch of $N$ input pairs (comprising $N-K$ ID pairs and $K$ OOD pairs), the latent representations are first obtained from the comment and code encoders. Then the margin-based loss is leveraged in the CL module to distinguish representations of ID and OOD data by constraining the cosine similarity. Specifically, the margin-based contrastive loss is first applied to $N-K$ ID code to maximize the difference between aligned (comment, code) pairs and incorrect pairs for each batch:

\begin{equation}\scriptsize
\label{eq:contrast_id}
    \hspace{-0.2cm} \mathcal{L}^{ID} = \sum_{i=1}^{N-K} \left(\frac{1}{N}\sum_{j=1, j \neq i}^N \max \left(0,m - s(v_{t_i^+}, v_{c_i^+}) + s(v_{t_j^-}, v_{c_i^+})\right)\right)
\end{equation}

\noindent $s(v_{t_i^+}, v_{c_i^+})$ represents the cosine similarity of representations between each aligned ID pair from all the $N-K$ aligned pairs, and $s(v_{t_j^-},v_{c_i^+})$ represents the cosine similarity of representations between each ID code and all the other $N-1$ comments (\ie the comment is either not aligned with the ID code or from OOD comments). Thus, this margin-based loss encourages the difference between the aligned pairs and the incorrect pairs greater than margin $m$. 

Regarding the $K$ OOD code, we enforce a constraint on the cosine similarity between each OOD code and all the comments, ensuring that the similarity remains below a margin $m$. This constraint is necessary because each OOD code should not align with its corresponding comment, nor with any of the other $K-1$ OOD comments and the $N-K$ ID comments. The loss function is denoted as follows:
\begin{equation}\small
\label{eq:contrast_ood}
\mathcal{L}^{OOD} = \sum_{k=1}^{K}\left(\frac{1}{N}\sum_{i=1}^N \max \left(0,-m + sim(t_j^-, c_k^-)\right)\right),
\end{equation}

\noindent where $sim(t_j^-, c_k^-)$ represents the cosine similarity between each of the $K$ OOD code and all $N$ comments.
Finally, the overall loss for the contrastive module can be expressed as:

\begin{equation}\small
\label{eq:contrast}
    \mathcal{L}^{CL} = \frac{1}{N}\left(\mathcal{L}^{ID}+ \mathcal{L}^{OOD}\right).
\end{equation}

\textbf{Binary OOD Rejection (BC) Module.}
Besides the CL module, we also introduce a classification module under weakly-supervision for identifying OOD samples. 
Inspired by the Replaced Token Detection (RTD) objective utilized in~\cite{feng2020codebert}, we bypass the generation phase since our OOD data are generated prior to training. Therefore, we directly train a rejection network responsible for determining whether (comment, code) pairs are OOD or not, which can be framed as a binary classification problem. Our binary OOD rejection network comprises a 3-layer fully-connected neural network with \textit{Tanh} activation, and the input is based on the concatenation of features from the comment and code encoders: 
$v_i = (v_{t_i}, v_{c_i}, v_{t_i} - v_{c_i}, v_{t_i} + v_{c_i}).$ Apart from utilizing the comment and code features, we also incorporate feature subtraction $v_{t_i} - v_{c_i}$ and aggregation $v_{t_i} + v_{c_i}$.  Additionally, we apply the sigmoid function to the output layer, producing a prediction probability that indicates whether the sample is OOD. We then use binary cross entropy loss for this module:

\begin{equation}\small
    \mathcal{L}^{BC} = \frac{1}{N}\sum_{i=1}^{N-K}(y_i\log p(v_i)+(1-y_i)\log(1-p(v_i))),
\end{equation}

\noindent where $p(v_i)$ is the output probability of the BC module, and $y_i \in [0, 1]$ is the ground-truth label. $y_i = 1$ indicates the input sample is an inlier, while $y_i = 0$ signifies it is an outlier. 

Hence, for weakly-supervised COOD+, we combine the objectives of the CL and the BC modules to jointly train our model, where $\lambda$ is a weight used to balance the loss functions:

\begin{equation}\small
\label{eq:loss}
    \mathcal{L} = \mathcal{L}^{CL} + \lambda \mathcal{L}^{BC}.
\end{equation}

\noindent\textbf{Combined Scoring Function.}
Similar to the unsupervised COOD approach, we utilize the diagonals of the similarity matrix as the OOD detection scores obtained from the CL module. To further improve the detection performance of the weakly-supervised version, we combine these $P^{CL}$ scores with the output probabilities of the BC module, denoted as $P^{BC}$. Here, we convert cosine similarity scores into probabilities using the sigmoid function $P^{CL*}=\sigma(sim(v_{c},v_{t}))$, then use multiplication to create the overall scoring function, yielding $P^{ID}=P^{CL*}\times P^{BC}$. We anticipate that higher scores will be assigned to ID pairs, while lower scores will be assigned to OOD pairs. This combined scoring function aims to enhance the discrimination between inliers and outliers, leading to more effective OOD detection.
\section{Empirical Evaluation Design} \label{sec:design}
To evaluate the performance of the proposed approaches in four scenarios, we investigate the following research questions:

\begin{enumerate}[label=\textbf{RQ$_\arabic*$:}, ref=\textbf{RQ$_\arabic*$}, wide, labelindent=5pt]\setlength{\itemsep}{0.2em}
    \item \label{rq:baseline}{\textit{How effective is our unsupervised COOD when compared to unsupervised baselines?}}
    \item \label{rq:athena}{\textit{How effective is our weakly-supervised COOD+ when compared to weakly-supervised baselines?}}
    \item \label{rq:athena}{\textit{How effective is our weakly-supervised COOD+ when using different modules or encode backbone?}}
    \item \label{rq:impact}{\textit{Is the main task (Code Search) performance affected by our COOD/COOD+ auxiliary, and to what extent?}}
\end{enumerate}

\subsection{Datasets}
In our experiments, we rely on two benchmark datasets: CodeSearchNet (CSN)~\cite{guo2020graphcodebert} and TLCS~\cite{salza2022effectiveness}. CSN contains bimodal data points consisting of code paired with function-level NL descriptions (\ie first lines of documentation comments) in six PLs (\eg Python, Java) collected from GitHub repositories. While CSN was originally created for a specific downstream task (\ie code search), it has since been widely adopted by large (NL, PL) models~\cite{guo2020graphcodebert, guo2022unixcoder} for pre-training due to the informative nature of bimodal instances. Large NL-PL models are first pre-trained across \textit{all} six languages, and then further fine-tuned for a \textit{specific} PL for some downstream task to enhance performance. For code search, the goal is to retrieve the most relevant code given a NL query, where CSN is widely used to further fine-tune a PL-specific code search model~\cite{feng2020codebert}. 


Salza \etal \cite{salza2022effectiveness} used training samples from CSN for pre-training, and created a new dataset sourced from StackOverflow (SO) for fine-tuning the code search model, involving only \textit{three} PLs: Java, Python, and JavaScript. Specifically, they leverage SO user questions as search queries and accepted answers as retrieved code snippets, which differ from GitHub comments and the corresponding code in CSN. We refer to this new dataset as TLCS. Existing work~\cite{Yang2017msr, lotter2018so, Abdalkareem2017OnCR} investigated code clones between SO and GitHub, demonstrating there exists \textit{only} 1-3\% code reuse. Besides code, user questions in SO are typically formulated before code answers, without concrete knowledge of what code answers will be, and are mostly written by end-users. Conversely, in GitHub, method docstrings (\ie comments) are often written following code snippets, and are mostly written by developers. These distinctions cause performance shortfall when directly applying models trained on CSN to TLCS without further fine-tuning or transfer learning~\cite{guo2022unixcoder, huang2021cosqa, salza2022effectiveness}. 

\subsection{OOD Scenarios}



We design four distinct OOD scenarios using the datasets described above, with CSN-Java and CSN-Python as inliers due to their common use for the pre-training of code models. 

\noindent\textbf{Scenario 1: Out-domain.} Following existing ML work~\cite{hendrycks2020pretrained, podolskiy2021revisiting, zhou2021contrastive}, we create an out-of-domain setting by choosing OOD samples from a different dataset than the training data. Thus, samples from TLCS-Java or TLCS-Python are treated as outliers accordingly. Inliers and their corresponding outliers belong to the same PL to ensure approaches don't identify OODs based on syntax differences between PLs but on data domains: GitHub vs. SO. Prior studies~\cite{wang2022enriching} show that CSN queries are longer than SO questions on average, so we sampled TLCS questions and answers to match the length distribution of CSN comments and code, to avoid OOD approaches exploiting spurious cues of query length differences. 
We didn't consider other code search datasets~\cite{yao2018staqc, li2020hierarchical, rao2021search4code} because they either contain only one of the PLs (Python or Java) or have a smaller dataset size. 

\noindent\textbf{Scenario 2: Misaligned.} In this scenario, we shuffle normal NL-PL pairs so that each code doesn't match its NL description. Although the NL modality sourced from attached comments in code are typically aligned with the PL modality, documentation errors may still occur and not effectively filtered by handcrafted rules~\cite{guo2020graphcodebert}. 

\noindent\textbf{Scenario 3: Shuffled-comment.} For (comment, code) pairs, we modify the syntactic information in each comment by shuffling 20\% of selected tokens using a seeded random algorithm~\cite{sinha-etal-2021-masked} with positions of stopwords and punctuations unchanged. No changes are made to the code for this scenario. This scenario is inspired by~\cite{sinha2020unnatural, mai2022self}. \cite{sinha2020unnatural} discovered that NL pre-trained models are insensitive to permuted sentences, which contrasts with human behavior as humans struggle to understand ungrammatical sentences, or interpret a completely different meaning from a few changes in word order. \cite{mai2022self} further introduces syntactic (shuffling) outliers into NL pre-training corpora to enhance OOD robustness and NL understanding performance.

\noindent\textbf{Scenario 4: Buggy-code.} We create buggy code using a \textit{semantic} mutation algorithm which injects more natural and realistic bugs into code than other traditional loose/strict mutators~\cite{richter2022learning}. This simulates buggy programs that the model may encounter during testing, typically absent from the training dataset, and should be taken into account by OOD code detectors according to the OOD definition~\cite{he2022distribution}. We avoid using real bug/vulnerability datasets~\cite{just2014defects4j, zhou2019devign, widyasari2020bugsinpy} due to limitations like the absence of paired comments, lack of support for Python or Java, introduction to a new dataset domain \etc. We generate buggy code for each code in CSN-Java and CSN-Python using~\cite{richter2022learning} to serve as outliers, ensuring the inliers and outliers are from the same dataset domain with the only difference being normal vs. buggy code. We focus on variable-misuse bugs, as only this mutation algorithm is available for both Python and Java in ~\cite{richter2022learning}. Variable-misuses occur when a variable name is used but another was meant in scope, and often remain undetected after compilation and regarded as hard-to-detect by recent bug detection techniques~\cite{he2022distribution, Richter2023how}. Comments remain unchanged for this scenario. 



\subsection{Model Configurations} 
For the weakly-supervised COOD+, we experiment with either the contrastive learning module (COOD+\_CL) or the binary OOD rejection module (COOD+\_BC) to compare against the combined model.  All models are trained using the Adam optimizer with a learning rate of $1e-5$, and a linear schedule with 10\% warmup steps. The batch size is set to 64, and the number of training epochs is 10. For the COOD+\_CL and COOD+, the margins in the margin-based loss are set to $0.2$. The balancing value $\lambda$ is set to $0.2$ after a grid search. The hidden layer size in the binary OOD rejection module for COOD+ is 384 ($768/2$). We also explore the robustness
and agnosticism of our COOD+ approach to different NL-PL models by replacing the GraphCodeBERT encoder with CodeBERT~\cite{feng2020codebert}, UniXcoder~\cite{guo2022unixcoder}, and ContraBERT~\cite{liu2023contrabert}.

\subsection{OOD detection model training and measurement}

For unsupervised COOD, we use only ID data for model training, thus involving all training data from CSN-Python and CSN-Java, with 10\% randomly sampled for validation. We avoid using the CSN development dataset for validation due to its smaller size. For weakly-supervised COOD+, we randomly select 1\% of the training data and replaced them with OOD samples generated for each scenario (following~\cite{tian2020few}), resulting in a total of 4\% OOD samples and 96\% ID samples for training. During inference, both COOD and COOD+ utilize the same ratio (20\%) for inliers and outliers from each scenario, which is more convincing than using an imbalanced dataset (\ie tiny number of OOD data). Detailed dataset statistics are provided in our online appendix~\cite{cood-tool}. Since all outliers are randomly selected, we report average OOD detection results across \textit{five} random seeds of the test dataset to ensure evaluation reliability and reproducibility.  
 
Following prior work in ML~\cite{hendrycks2019using, liznerski2022exposing}, we use two standard metrics to measure the effectiveness of our COOD/COOD+ models: the area under the receiver operating characteristic curve (AUROC) and the false positive rate at 95\% (FPR95). AUROC is threshold-independent, calculating the area under the ROC curve over a range of threshold values, representing the trade-off between true positive rate and false positive rate. It quantifies the probability that a positive example (ID sample) receives a higher score than a negative one (OOD sample). Higher AUROC indicates better performance. Additionally, FPR95 corresponds to the false positive rate (FPR) when the true positive rate of ID samples is 95\%. FPR95 is threshold-dependent, where OODs are identified by setting a threshold $\sigma$ with $ P^{OOD}<1-\sigma$ ($P^{ID}>\sigma$) so that a high fraction (95\%) of ID data is above the threshold. It measures the proportion of OOD samples that are mistakenly classified when 95\% of ID samples are correctly recalled based on the threshold. Lower FPR95 indicates better performance.

\subsection{Baselines}
We compare our COOD/COOD+ against various OOD detection baselines, including adaptations of existing unsupervised NLP OOD approaches on NL-PL encoders (1-2), weakly-supervised approaches based on outlier exposure (3), and neural bug detection techniques (4-5). Since unsupervised approaches (1-2) rely on classification outputs for OOD scoring, we reformulate code search as binary classification to fine-tune the encoders similarly to~\cite{feng2020codebert}. (1-2) is supervised for code search, but unsupervised for OOD detection. For weakly-supervised baselines (3-5), we use the same number of OOD samples as COOD+ for a fair comparison. Note that the encoder backbone of (1-3) is also GraphCodeBERT. (4-5) are specifically designed for neural bug detection, thus not requiring other encoder backbone for OOD detection. 

\begin{enumerate}[leftmargin=1.5em] 
    \item \textbf{Supervised Contrastive Learning For Classification (SCL)}~\cite{khosla2020supervised}. This method fine-tunes transformer-based classification models by maximizing similarity of input pairs if they are from the same class and minimize it otherwise. Following \cite{zhou2021contrastive}, we adopts MSP, Energy, and Mahalanobis OOD scoring algorithms for OOD detection.
    
    \item \textbf{Margin-based Contrastive Learning for Classification (MCL)}~\cite{zhou2021contrastive}. This approach fine-tunes transformer-based classification models by minimizing the L2 distances between instances from the same class, and encouraging the L2 distances between instances of different classes to exceed a margin. We also detect OODs by applying MSP, Energy, and Mahalanobis OOD scoring algorithms.
    
    \item \textbf{Energy-based Outlier Exposure (EOE)}~\cite{liu2020energy}. This approach uses a few auxiliary OOD data to fine-tune the classification model with an energy-based margin loss~\cite{liu2020energy}, and then utilize Energy scores for OOD detection. 
    
    \item \textbf{CuBERT}~\cite{kanade2020learning}. CuBERT is pre-trained on a large code corpus using masked language modeling, then fine-tuned for bug detection and repair. We adapt CuBERT for OOD classification by alternatively fine-tuning it on our datasets with comments appended to their corresponding code, as CuBERT only accepts single instance inputs.

    \item \textbf{2P-CuBERT}~\cite{he2022distribution}. This method enhances CuBERT's bug detection accuracy with a two-phase fine-tuning approach. The first phase utilizes contrastive learning on generated synthetic buggy code~\cite{allamanis2021self}. For the second phase, we alternatively fine-tune CuBERT to detect OOD using our datasets. Results are reported only for CSN-Python due to the lack of Java bug generation algorithms in~\cite{he2022distribution}.

\end{enumerate}

\subsection{Main Task Performance Analysis}
An effective OOD detector, serving as an auxiliary component, should identify and reject OOD samples without negatively impacting the original model's performance on the main downstream task with ID data~\cite{zhou2021contrastive}. Consequently, we validate the effectiveness of our COOD/COOD+ auxiliary on the code search task using the official evaluation benchmark~\cite{guo2020graphcodebert, liu2023contrabert} by calculating the mean reciprocal rank (mRR) for each pair of comment-code data over distractor codes in the testing code corpus. Specifically, we first measure the performance of original GraphCodeBERT code search model on both ID and OOD data, whose performance is expected to be negatively affected with the presence of OOD samples. Then, we utilize our COOD/COOD+ auxiliary to filter the testing dataset by setting a threshold to retain 95\% of ID instances with higher scores (following existing ML work~\cite{ming2022delving} and the FPR95 definition), as real-world deployment typically involves few OODs. Finally, we directly use the fine-tuned encoder in COOD/COOD+ to perform code search but on the retained ID instances, and compare this performance with that on the ground-truth ID instances. If the performance loss is recovered by using COOD/COOD+, we actually enhance the trustworthiness and robustness of the original code search model (as shown in Sec. VI-D). Here trustworthiness and robustness mean that predictions of code models become more reliable when encountering OOD data in real-world deployment. Note that the dataset used for COOD/COOD+ training is the same as that used for PL-specific training of existing SOTA code search models.

\begin{table*}[thb]
\centering
\footnotesize
\caption{Effectiveness of our COOD and COOD+ models compared with the baselines on the CSN-Python dataset.}
\label{tab:results-python}
\begin{adjustbox}{width=0.9\textwidth}\small
\begin{tabular}{|ccccccccccc|}
\hline
\multicolumn{1}{|c|}{\multirow{2}{*}{\textbf{Approaches}}} & \multicolumn{2}{c|}{\textbf{Out-domain+ID}}                                 & \multicolumn{2}{c|}{\textbf{Misaligned+ID}}                                 & \multicolumn{2}{c|}{\textbf{Shuffled-comment+ ID}}                          & \multicolumn{2}{c|}{\textbf{Buggy-code+ ID}}                                & \multicolumn{2}{c|}{\textbf{Overall (All OODs+ID)}}    \\ \cline{2-11} 
\multicolumn{1}{|c|}{}                            & \multicolumn{1}{c|}{\textbf{AUROC↑}} & \multicolumn{1}{c|}{\textbf{FPR95↓}} & \multicolumn{1}{c|}{\textbf{AUROC↑}} & \multicolumn{1}{c|}{\textbf{FPR95↓}} & \multicolumn{1}{c|}{\textbf{AUROC↑}} & \multicolumn{1}{c|}{\textbf{FPR95↓}} & \multicolumn{1}{c|}{\textbf{AUROC↑}} & \multicolumn{1}{c|}{\textbf{FPR95↓}} & \multicolumn{1}{c|}{\textbf{AUROC↑}} & \textbf{FPR95↓} \\ \hline
\multicolumn{11}{|c|}{\textbf{Unsupervised}}                                                                                                                                                                                                                                                                                                                                                                                \\ \hline
\multicolumn{1}{|c|}{SCL+MSP}              & \multicolumn{1}{c|}{78.21}           & \multicolumn{1}{c|}{68.93}           & \multicolumn{1}{c|}{49.80}           & \multicolumn{1}{c|}{97.23}           & \multicolumn{1}{c|}{60.91}           & \multicolumn{1}{c|}{90.36}           & \multicolumn{1}{c|}{49.23}           & \multicolumn{1}{c|}{95.26}           & \multicolumn{1}{c|}{60.79}           & 87.05           \\ \hline
\multicolumn{1}{|c|}{SCL+Energy}           & \multicolumn{1}{c|}{77.76}           & \multicolumn{1}{c|}{70.13}           & \multicolumn{1}{c|}{65.07}           & \multicolumn{1}{c|}{96.11}           & \multicolumn{1}{c|}{61.24}           & \multicolumn{1}{c|}{90.04}           & \multicolumn{1}{c|}{49.58}           & \multicolumn{1}{c|}{95.00}           & \multicolumn{1}{c|}{65.09}           & 86.95           \\ \hline
\multicolumn{1}{|c|}{SCL+Maha}             & \multicolumn{1}{c|}{73.34}           & \multicolumn{1}{c|}{82.82}           & \multicolumn{1}{c|}{73.21}           & \multicolumn{1}{c|}{92.13}           & \multicolumn{1}{c|}{68.03}           & \multicolumn{1}{c|}{87.02}           & \multicolumn{1}{c|}{\textbf{53.89}}  & \multicolumn{1}{c|}{\textbf{92.43}}  & \multicolumn{1}{c|}{68.72}           & 88.14           \\ \hline
\multicolumn{1}{|c|}{MCL+MSP}              & \multicolumn{1}{c|}{81.18}           & \multicolumn{1}{c|}{65.57}           & \multicolumn{1}{c|}{53.51}           & \multicolumn{1}{c|}{96.43}           & \multicolumn{1}{c|}{62.17}           & \multicolumn{1}{c|}{90.44}           & \multicolumn{1}{c|}{48.69}           & \multicolumn{1}{c|}{94.42}           & \multicolumn{1}{c|}{63.11}           & 85.78           \\ \hline
\multicolumn{1}{|c|}{MCL+Energy}           & \multicolumn{1}{c|}{82.55}           & \multicolumn{1}{c|}{64.03}           & \multicolumn{1}{c|}{62.57}           & \multicolumn{1}{c|}{95.43}           & \multicolumn{1}{c|}{63.03}           & \multicolumn{1}{c|}{90.52}           & \multicolumn{1}{c|}{\textbf{48.26}}  & \multicolumn{1}{c|}{94.90}           & \multicolumn{1}{c|}{66.02}           & 85.16           \\ \hline
\multicolumn{1}{|c|}{MCL+Maha}             & \multicolumn{1}{c|}{53.05}           & \multicolumn{1}{c|}{94.60}           & \multicolumn{1}{c|}{62.47}           & \multicolumn{1}{c|}{93.12}           & \multicolumn{1}{c|}{49.23}           & \multicolumn{1}{c|}{95.73}           & \multicolumn{1}{c|}{51.13}           & \multicolumn{1}{c|}{93.64}           & \multicolumn{1}{c|}{54.31}           & 94.35           \\ \hline
\multicolumn{1}{|c|}{COOD}                 & \multicolumn{1}{c|}{\textbf{86.60}}  & \multicolumn{1}{c|}{\textbf{48.25}}  & \multicolumn{1}{c|}{\textbf{99.85}}  & \multicolumn{1}{c|}{\textbf{0.16}}   & \multicolumn{1}{c|}{\textbf{72.82}}  & \multicolumn{1}{c|}{\textbf{85.18}}  & \multicolumn{1}{c|}{49.17}           & \multicolumn{1}{c|}{\textbf{93.58}}  & \multicolumn{1}{c|}{\textbf{80.50}}  & \textbf{52.31}  \\ \hline
\multicolumn{11}{|c|}{\textbf{Weakly-supervised}}                                                                                                                                                                                                                                                                                                                                                                           \\ \hline
\multicolumn{1}{|c|}{EOE}                  & \multicolumn{1}{c|}{\textbf{98.96}}  & \multicolumn{1}{c|}{\textbf{3.26}}   & \multicolumn{1}{c|}{91.18}           & \multicolumn{1}{c|}{50.03}           & \multicolumn{1}{c|}{\textbf{98.37}}  & \multicolumn{1}{c|}{\textbf{4.07}}   & \multicolumn{1}{c|}{94.78}  & \multicolumn{1}{c|}{24.69}           & \multicolumn{1}{c|}{95.95}           & 20.02           \\ \hline
\multicolumn{1}{|c|}{CuBERT}               & \multicolumn{1}{c|}{92.56}           & \multicolumn{1}{c|}{14.46}           & \multicolumn{1}{c|}{91.13}           & \multicolumn{1}{c|}{17.33}           & \multicolumn{1}{c|}{88.92}           & \multicolumn{1}{c|}{21.74}           & \multicolumn{1}{c|}{60.93}           & \multicolumn{1}{c|}{77.73}           & \multicolumn{1}{c|}{86.11}           & 27.38           \\ \hline
\multicolumn{1}{|c|}{2P-CuBERT}            & \multicolumn{1}{c|}{92.26}           & \multicolumn{1}{c|}{15.16}           & \multicolumn{1}{c|}{84.42}  & \multicolumn{1}{c|}{30.83}  & \multicolumn{1}{c|}{86.38}           & \multicolumn{1}{c|}{26.92}           & \multicolumn{1}{c|}{92.87}           & \multicolumn{1}{c|}{13.94}  & \multicolumn{1}{c|}{88.51}  & 22.65  \\ \hline
\multicolumn{1}{|c|}{COOD+}                & \multicolumn{1}{c|}{98.80}           & \multicolumn{1}{c|}{4.30}            & \multicolumn{1}{c|}{\textbf{99.53}}  & \multicolumn{1}{c|}{\textbf{0.40}}   & \multicolumn{1}{c|}{98.02}           & \multicolumn{1}{c|}{6.52}            & \multicolumn{1}{c|}{\textbf{97.90}}  & \multicolumn{1}{c|}{\textbf{5.96}}   & \multicolumn{1}{c|}{\textbf{98.64}}  & \textbf{4.09}   \\
\multicolumn{1}{|c|}{COOD+\_CL}            & \multicolumn{1}{c|}{93.91}           & \multicolumn{1}{c|}{25.44}           & \multicolumn{1}{c|}{99.89}           & \multicolumn{1}{c|}{0.17}            & \multicolumn{1}{c|}{82.57}           & \multicolumn{1}{c|}{72.97}           & \multicolumn{1}{c|}{52.56}           & \multicolumn{1}{c|}{93.99}           & \multicolumn{1}{c|}{85.83}           & 42.57           \\
\multicolumn{1}{|c|}{COOD+\_BC}            & \multicolumn{1}{c|}{96.49}           & \multicolumn{1}{c|}{8.67}            & \multicolumn{1}{c|}{74.27}           & \multicolumn{1}{c|}{61.38}           & \multicolumn{1}{c|}{97.53}           & \multicolumn{1}{c|}{5.68}            & \multicolumn{1}{c|}{95.71}           & \multicolumn{1}{c|}{10.95}           & \multicolumn{1}{c|}{90.43}           & 22.98           \\ \hline

\end{tabular}
\end{adjustbox}
\end{table*}

\begin{table*}[]
\centering
\footnotesize
\caption{Effectiveness of our COOD and COOD+ models compared with the baselines on the CSN-Java dataset.}
\label{tab:results-java}
\begin{adjustbox}{width=0.9\textwidth}\small
\begin{tabular}{|ccccccccccc|}
\hline
\multicolumn{1}{|c|}{\multirow{2}{*}{\textbf{Approaches}}} & \multicolumn{2}{c|}{\textbf{Out-domain+ID}}                                 & \multicolumn{2}{c|}{\textbf{Misaligned+ID}}                                 & \multicolumn{2}{c|}{\textbf{Shuffled-comment+ ID}}                          & \multicolumn{2}{c|}{\textbf{Buggy-code+ ID}}                                & \multicolumn{2}{c|}{\textbf{Overall (All OODs+ID)}}    \\ \cline{2-11} 
\multicolumn{1}{|c|}{}                                     & \multicolumn{1}{c|}{\textbf{AUROC↑}} & \multicolumn{1}{c|}{\textbf{FPR95↓}} & \multicolumn{1}{c|}{\textbf{AUROC↑}} & \multicolumn{1}{c|}{\textbf{FPR95↓}} & \multicolumn{1}{c|}{\textbf{AUROC↑}} & \multicolumn{1}{c|}{\textbf{FPR95↓}} & \multicolumn{1}{c|}{\textbf{AUROC↑}} & \multicolumn{1}{c|}{\textbf{FPR95↓}} & \multicolumn{1}{c|}{\textbf{AUROC↑}} & \textbf{FPR95↓} \\ \hline
\multicolumn{11}{|c|}{\textbf{Unsupervised}}                                                                                                                                                                                                                                                                                                                                                                              \\ \hline
\multicolumn{1}{|c|}{SCL+MSP}            & \multicolumn{1}{c|}{85.15}           & \multicolumn{1}{c|}{63.66}           & \multicolumn{1}{c|}{58.93}           & \multicolumn{1}{c|}{95.76}           & \multicolumn{1}{c|}{58.51}           & \multicolumn{1}{c|}{92.59}           & \multicolumn{1}{c|}{49.01}           & \multicolumn{1}{c|}{95.70}           & \multicolumn{1}{c|}{62.91}           & 86.92           \\ \hline
\multicolumn{1}{|c|}{SCL+Energy}         & \multicolumn{1}{c|}{84.00}           & \multicolumn{1}{c|}{66.69}           & \multicolumn{1}{c|}{55.27}           & \multicolumn{1}{c|}{95.56}           & \multicolumn{1}{c|}{60.07}           & \multicolumn{1}{c|}{91.82}           & \multicolumn{1}{c|}{49.13}           & \multicolumn{1}{c|}{95.86}           & \multicolumn{1}{c|}{62.12}           & 87.48           \\ \hline
\multicolumn{1}{|c|}{SCL+Maha}           & \multicolumn{1}{c|}{82.16}           & \multicolumn{1}{c|}{73.57}           & \multicolumn{1}{c|}{79.20}           & \multicolumn{1}{c|}{89.71}           & \multicolumn{1}{c|}{64.61}           & \multicolumn{1}{c|}{91.68}           & \multicolumn{1}{c|}{46.55}           & \multicolumn{1}{c|}{97.32}           & \multicolumn{1}{c|}{68.13}           & 88.07           \\ \hline
\multicolumn{1}{|c|}{MCL+MSP}            & \multicolumn{1}{c|}{85.16}           & \multicolumn{1}{c|}{65.74}           & \multicolumn{1}{c|}{59.12}           & \multicolumn{1}{c|}{95.83}           & \multicolumn{1}{c|}{59.50}           & \multicolumn{1}{c|}{95.73}           & \multicolumn{1}{c|}{49.44}           & \multicolumn{1}{c|}{95.73}           & \multicolumn{1}{c|}{63.31}           & 87.40           \\ \hline
\multicolumn{1}{|c|}{MCL+Energy}         & \multicolumn{1}{c|}{83.44}           & \multicolumn{1}{c|}{68.53}           & \multicolumn{1}{c|}{44.85}           & \multicolumn{1}{c|}{96.22}           & \multicolumn{1}{c|}{59.26}           & \multicolumn{1}{c|}{91.79}           & \multicolumn{1}{c|}{\textbf{49.61}}  & \multicolumn{1}{c|}{95.51}           & \multicolumn{1}{c|}{59.45}           & 88.01           \\ \hline
\multicolumn{1}{|c|}{MCL+Maha}           & \multicolumn{1}{c|}{50.36}           & \multicolumn{1}{c|}{96.73}           & \multicolumn{1}{c|}{67.11}           & \multicolumn{1}{c|}{90.61}           & \multicolumn{1}{c|}{48.44}           & \multicolumn{1}{c|}{96.07}           & \multicolumn{1}{c|}{46.68}           & \multicolumn{1}{c|}{97.44}           & \multicolumn{1}{c|}{53.00}           & 95.21           \\ \hline
\multicolumn{1}{|c|}{COOD}               & \multicolumn{1}{c|}{\textbf{92.27}}  & \multicolumn{1}{c|}{\textbf{40.14}}  & \multicolumn{1}{c|}{\textbf{99.41}}  & \multicolumn{1}{c|}{\textbf{0.39}}   & \multicolumn{1}{c|}{\textbf{75.78}}  & \multicolumn{1}{c|}{\textbf{86.88}}  & \multicolumn{1}{c|}{48.72}           & \multicolumn{1}{c|}{\textbf{95.04}}  & \multicolumn{1}{c|}{\textbf{79.05}}  & \textbf{55.95}  \\ \hline
\multicolumn{11}{|c|}{\textbf{Weakly-supervised}}                                                                                                                                                                                                                                                                                                                                                                         \\ \hline
\multicolumn{1}{|c|}{EOE}                & \multicolumn{1}{c|}{\textbf{99.49}}  & \multicolumn{1}{c|}{\textbf{1.59}}   & \multicolumn{1}{c|}{88.83}           & \multicolumn{1}{c|}{62.12}           & \multicolumn{1}{c|}{\textbf{98.71}}  & \multicolumn{1}{c|}{\textbf{3.58}}   & \multicolumn{1}{c|}{\textbf{92.27}}  & \multicolumn{1}{c|}{25.51}           & \multicolumn{1}{c|}{94.82}           & 23.22           \\ \hline
\multicolumn{1}{|c|}{CuBERT}             & \multicolumn{1}{c|}{82.59}           & \multicolumn{1}{c|}{62.23}           & \multicolumn{1}{c|}{49.38}           & \multicolumn{1}{c|}{95.25}           & \multicolumn{1}{c|}{50.14}           & \multicolumn{1}{c|}{94.93}           & \multicolumn{1}{c|}{67.97}           & \multicolumn{1}{c|}{94.82}           & \multicolumn{1}{c|}{62.53}           & 86.80           \\ \hline
\multicolumn{1}{|c|}{COOD+}              & \multicolumn{1}{c|}{99.29}           & \multicolumn{1}{c|}{2.79}            & \multicolumn{1}{c|}{\textbf{99.56}}  & \multicolumn{1}{c|}{\textbf{0.81}}   & \multicolumn{1}{c|}{97.05}           & \multicolumn{1}{c|}{9.43}            & \multicolumn{1}{c|}{91.11}           & \multicolumn{1}{c|}{\textbf{20.88}}  & \multicolumn{1}{c|}{\textbf{96.75}}  & \textbf{8.48}   \\
\multicolumn{1}{|c|}{COOD+\_CL}          & \multicolumn{1}{c|}{93.73}           & \multicolumn{1}{c|}{23.76}           & \multicolumn{1}{c|}{99.65}           & \multicolumn{1}{c|}{0.31}            & \multicolumn{1}{c|}{83.24}           & \multicolumn{1}{c|}{78.14}           & \multicolumn{1}{c|}{50.12}           & \multicolumn{1}{c|}{95.53}           & \multicolumn{1}{c|}{82.47}           & 47.96           \\
\multicolumn{1}{|c|}{COOD+\_BC}          & \multicolumn{1}{c|}{98.67}           & \multicolumn{1}{c|}{3.92}            & \multicolumn{1}{c|}{64.55}           & \multicolumn{1}{c|}{78.89}           & \multicolumn{1}{c|}{95.35}           & \multicolumn{1}{c|}{10.17}           & \multicolumn{1}{c|}{94.09}           & \multicolumn{1}{c|}{14.40}           & \multicolumn{1}{c|}{88.16}           & 26.86           \\ \hline

\end{tabular}
\end{adjustbox}
\end{table*}

\section{Experimental Results}\label{sec:results}
\subsection{RQ1: Unsupervised COOD Performance}
In this subsection, we analyze the experimental results to assess the detection performance of our unsupervised COOD model compared with the unsupervised baselines. According to Table~\ref{tab:results-python} and~\ref{tab:results-java}, we can observe that COOD outperforms all unsupervised baselines on both CSN-Python and CSN-Java. Notably, COOD effectively detect \textit{out-domain} and \textit{misaligned OOD} testing samples, while other unsupervised approaches only work for the \textit{out-domain} scenario. This is because COOD effectively captures alignment information within (comment, code) pairs through a multi-modal contrastive learning objective with InfoNCE loss and uses similarity scores between comments and code to detect OODs. Specifically, COOD outputs low similarity scores for the out-domain data from TLCS by additionally considering the knowledge gap difference in (comment, code) pairs between ID and out-domain data. Also, as the misaligned scenario involves misaligned (comment, code) pairs, their similarity scores are naturally low. In contrast, the unsupervised baselines aggregate misaligned information into classification logits and rely on the confidence of the "aligned" class to detect OODs. As previously discussed in Sec. IV-C, the contrastive losses~\cite{khosla2020supervised, zhou2021contrastive} used by them are not as effective for learning alignment information, leading to inferior performance. Additionally, detecting token-level OOD in \textit{shuffled-comment} and \textit{buggy-code} scenarios proves challenging without seeing OOD samples during training, as all unsupervised methods fail to detect these OODs. 



\subsection{RQ2: Weakly-supervised COOD+ Performance}
We further investigate the performance of our weakly-supervised COOD+ method against several weakly-supervised baselines on CSN-Python and CSN-Java. Table~\ref{tab:results-python} shows that weak supervision on a tiny amount of OOD data enables COOD+ (and EOE) to not only address unsupervised COOD's shortcomings in detecting finer-grained \textit{shuffled-comment} and \textit{buggy-code} OODs, but also enhance performance for the \textit{out-domain} scenario for CSN-Python. This improvement aligns with previous research ~\cite{hendrycks2018deep, liu2020energy, kim-etal-2023-pseudo} which enhances OOD detection by complementing the downstream task objective with an complementary discriminator operating to distinguish IDs from external OODs. While EOE slightly outperforms COOD+ for the \textit{out-domain} and \textit{shuffled-comment} scenarios by utilizing the prediction probabilities from one classification module, our COOD+, which combines the BC and CL modules, delivers consistently high performance across all four scenarios, resulting in superior overall performance. In addition, the BC module can be directly adapted to the overall COOD+ framework without modifying the underlying learning objective, but the outlier exposure-based methods (\eg EOE) typically require additional engineering (\eg determining class-probabilistic distributions~\cite{hendrycks2018deep}, boundaries for energy scores~\cite{liu2020energy}) to equip ML models with OOD detection abilities. Besides, the bug detection method 2P-CuBERT can reasonably detect OODs, but its performance for the \textit{buggy-code} scenario is negatively impacted by the limited amount of training OOD examples.

On the CSN-Java dataset, our COOD+ also achieves the best overall performance compared to all baselines, despite trailing slightly behind EOE for \textit{out-domain} and \textit{shuffled-comment} OODs. While EOE has higher AUROC score than that of COOD+ for the \textit{buggy-code} scenario, it suffers from a high FPR95, indicating a higher margin of error for OOD inference using a threshold of 95\% ID recall. Moreover, similar to CSN-Python, CuBERT fails to detect OODs effectively on CSN-Java either, likely due to the lack of training examples. In summary, the superior performance of our COOD+ model results from the interplay between the CL and BL modules, where contrastive learning captures high-level alignment between NL-PL input pairs that is naturally suitable for \textit{out-domain} and \textit{misaligned} OODs, while the OOD rejection classifier targets lower-level OOD information from \textit{shuffled-comment} and \textit{buggy-code} samples. Furthermore, by utilizing a weakly-supervised contrastive learning objective that jointly optimizes for OOD detection and the code search task, our method enables effective deployment of the code search model in OOD environments, which will be further studied in Sec. VI-D.

\subsection{RQ3: Weakly-Supervised COOD+ Performance with Different Model Components and Encoder Backbone}
In this subsection, we evaluate the effect of using only the CL (COOD+\_CL) or the BC module (COOD+\_BC) against the proposed combined COOD+ model to illustrate how COOD+ generalizes in four OOD scenarios. As shown in Table~\ref{tab:results-python} and~\ref{tab:results-java}, COOD+\_CL performs well in the \textit{out-domain} and \textit{misaligned} scenarios, which is due to its ability to effectively capture high-level (comment, code) alignment information. COOD+\_BC excels in the \textit{out-domain}, \textit{shuffled-comment}, and \textit{buggy-code} scenarios, since it can learn lower-level features from these types of OOD samples. While COOD+\_BC maintains acceptable OOD detection performance with high AUROC ($>$90\%) and low FPR95 ($<$25\%), the CL module remains crucial for overall performance, since without it the overall performance of COOD+ will drop below the EOE baseline. Moreover, removing the BC module has a more negative impact on the OOD detection as COOD+ loses the ability to capture the necessary lower-level OOD information for detecting \textit{shuffled-comment} and \textit{buggy-code} OODs. Note that the standalone CL module performs better than the unsupervised COOD overall, demonstrating that our proposed modification to the original CL objective enhance OOD detection by leveraging the margin-based loss. Thus, the combined model's superior performance validates our design choices. That is, the combined scoring function (cosine similarities from CL and the prediction probabilities from BC) is thoughtfully designed to leverage the advantage of each module for high detection accuracy.


\begin{table}[!tb]
\centering
\caption{Our COOD+ model with different encoders.}
\footnotesize
\label{tab:ablation}
\begin{adjustbox}{width=0.48\textwidth}\small
\begin{tabular}{c|cc|cc}
\toprule
\multirow{2}{*}{\textbf{Encoders}} & \multicolumn{2}{c|}{\textbf{CSN-Java}} & \multicolumn{2}{c}{\textbf{CSN-Python}} \\ \cline{2-5} 
                                   & \textbf{AUROC↑}    & \textbf{FPR95↓}   & \textbf{AUROC↑}    & \textbf{FPR95↓}    \\ \hline
GraphCodeBERT                      & \textbf{96.75}     & \textbf{8.48}     & \textbf{98.64}     & \textbf{4.09}               \\
CodeBERT                           & 95.42              & 10.27              & 98.59     & 4.20      \\
UniXcoder                          & 95.49              & 10.63              & 97.83              & 5.91               \\
ContraBERT                         & 96.19              & 9.32              & 98.25              & 4.76               \\ \bottomrule
\end{tabular}
\end{adjustbox}
\end{table}

Moreover, we compare the detection performance of our COOD+ with various underlying NL-PL pre-trained encoder. Specifically, we compare our choice of GraphCodeBERT~\cite{guo2020graphcodebert} against other NL-PL encoders from the literature including its predecessor, CodeBERT~\cite{feng2020codebert}, and more recent ones such as UniXcoder~\cite{guo2022unixcoder} and ContraBERT~\cite{liu2023contrabert}. As shown in Table~\ref{tab:ablation}, all encoders perform within a 1-2\% difference, indicating that our COOD+ framework is robust across different encoders. This demonstrates our framework's flexibility and effectiveness in detecting OODs when deploying various NL-PL encoders for code-related tasks. Furthermore, we investigate key hyperparameters in COOD+, such as $m$ for margin-based contrastive loss and  $\lambda$ in the overall loss function. The detailed results are available in our online appendix~\cite{cood-tool}.

\subsection{RQ4: Main Task Performance}
We present the code search performance under the impact of OOD instances by using GraphCodeBERT (GCB), COOD/COOD+, and the closest competitor EOE in Table~\ref{tab:maintask}. As described in Sec. V-F, we use the official metric mRR  and follow the same testing scheme as the original GraphCodeBERT code search model for evaluation. From Table~\ref{tab:maintask}, we first observe that our COOD/COOD+ achieves performance comparable to GraphCodeBERT, while the EOE suffers from a significant reduction in performance, as it reformulates code search as binary classification to gain OOD detection ability. This reveals a critical trade-off between OOD detection and downstream task performance. To further validate the importance of OOD detection for code search, we construct outliers based on the CSN-Java and -Python testing dataset, respectively. Given that code search aims to retrieve the most aligned code from a code corpus given an NL query, the outliers are only sampled from three OOD scenarios: \textit{out-domain}, \textit{shuffled-comment} and \textit{buggy-code}, each replacing 5\% ID data of the original testing set. We then show the results when the dataset contains 15\% OOD samples (\ie 15\% outliers), discard OOD samples by filtering the testing set by ground-truth labels (\ie Filtered-GT) or using various OOD detection models (\ie Filtered-OOD-model). Note that the Filtered-GT dataset is the original CSN's subset with 15\% of ID samples removed.

\begin{table}[!tb]
\centering
\caption{Code search performance under the impact of OOD detection. Higher numbers represent better performance}
\label{tab:maintask}
\begin{adjustbox}{width=0.48\textwidth}\small
\begin{tabular}{c|c|c|c|c|c}
\toprule
\textbf{Dataset}            & \textbf{Testing Subset} & \textbf{GCB} & \textbf{EOE} & \textbf{COOD} & \textbf{COOD+} \\ \hline
\multirow{4}{*}{CSN-Python} & Origin                  & 69.20        & 50.11        & 68.47         & 69.69          \\
                            & 15\% outliers            & 65.85        & 43.68        & 64.67         & 65.67          \\
                            & Filtered-GT  & 70.24        & 44.85        & 68.95         & 70.24          \\
                            & Filtered-OOD-model     & --           & 46.82        & 70.30         & \textbf{73.10} \\ \hline
\multirow{4}{*}{CSN-Java}   & Origin                  & 69.10        & 46.29        & 68.85         & 69.46          \\
                            & 15\% outliers            & 64.99        & 37.77        & 64.86         & 64.54          \\
                            & Filtered-GT  & 69.12        & 38.94        & 69.36         & 69.93          \\
                            & Filtered-OOD-model     & --           & 39.33        & 71.02         & \textbf{73.18} \\ \bottomrule
\end{tabular}
\end{adjustbox}
\end{table}

According to Table~\ref{tab:maintask}, the performance of the original GraphCodeBERT code search model drops by 4.84\% and 5.95\% ((69.10-64.99)/69.10) mRR when outliers are present in CSN-Python and -Java, respectively. As a solution to this issue, our COOD/COOD+ detector recover the performance losses by identifying and filtering out the OOD samples without negatively impacting the model's code understanding ability in code search. Specifically, the code search performance of COOD/COOD+ on the Filtered-COOD/COOD+ dataset (70.30\%/73.10\% and 71.02\%/73.18\% on CSN-Python and -Java, respectively) is comparable to or even better than GraphCodeBERT on the Filtered-GT dataset (70.24\% and 69.12\% on CSN-Python and -Java, respectively). This slight improvement is probably because our detectors filter out additional lower-quality testing samples that resemble outliers. Thus, our COOD/COOD+ enhance the trustworthiness and robustness of the GraphCodeBERT, since the model's predictions become more reliable when encountering OOD data. 
Note that the original GraphCodeBERT is not equipped with the OOD detection ability, so its corresponding cells for the Filtered-OOD-model in Table~\ref{tab:maintask} are left blank.

\section{Discussions}\label{sec:discussions}

\textbf{Analysis of the Overconfidence of MSP with Conformal Prediction.} DNN models pre-trained on ID data are prone to misclassify OOD samples as ID due to overconfident MSP scores~\cite{lee2017training, liang2017enhancing}. This issue arises from spurious correlations between OOD and ID features, such as entities or syntactic patterns in NL and PL data~\cite{zheng2020aug, wu2022revisit}. For example, OOD inputs with common ID syntactic patterns may receive high ID scores. To overcome overconfident predictions, previous work in ML explored techniques like temperature scaling~\cite{liang2017enhancing} and  confidence calibration using adversarial samples~\cite{lee2017training, bitterwolf2020certifiably}. In contrast, COOD+ leverages weakly-supervised contrastive learning with a small set of OOD samples to prevent the alignment of OOD pairs, and adopts a binary OOD rejection module to better differentiate OOD and ID representations. We further verify COOD+'s ability to address overconfidence using Conformal Prediction (CP)~\cite{angelopoulos2023conformal}. Post-hoc CP converts OOD scores into prediction sets with a high user-specified coverage (e.g., 95\%), correcting overconfident thresholds during calibration. This ensures prediction sets conform statistically to the desired coverage. We experimentally demonstrated that COOD+ achieves near-optimal prediction set sizes with 95\% coverage, effectively identifying true OODs with statistical guarantees. In contrast, MCL+MSP, the best MSP-based method, still struggles with overconfident OODs.  Therefore, COOD+ can effectively overcome the overconfidence issue of MSP. Full details of the experiments and result analysis are in the online appendix~\cite{cood-tool}.

\textbf{OOD detection with large language models (LLMs).} It's worth noting that transformer-based code models (\eg GraphCodeBERT~\cite{guo2020graphcodebert}) and LLMs share the same underlying transformer architecture. Scaling up transformer-based code models and training them on vast amounts of code data allows LLMs~\cite{roziere2023code} to perform a wide range of code-related tasks, making coding less labor-intensive and more accessible to end-users. Since LLMs are transformer-based, they are also vulnerable to OOD data, with potentially worse performance degradation due to error accumulation during auto-regressive inference. Thus, identifying OOD samples is crucial to knowing when to trust LLM outputs. Our proposed OOD code framework techniques can be applied to these larger transformer-based code models, similarly as demonstrated in our experiments with different encoders in Table~\ref{tab:ablation}.


\textbf{Generalization of COOD/COOD+ to other code-related tasks.} During software development, developers often write comments following code snippets (methods/functions). Therefore, from a realistic perspective, our framework is generalizable to many code understanding tasks beyond code search, such as clone detection and defect detection, that use (comment, code) pairs as input. All that is needed is identifying the ID dataset and out-domain data, as the four OOD scenarios outlined in this paper are relevant to most tasks. For instance, in clone detection, (comment, code) pairs could first be processed by our framework to identify OOD samples before checking for clones. Unfortunately, since existing clone and defect detection datasets typically lack comments, we haven't applied our framework to these tasks. However, the framework has strong potential to enhance these tasks as more realistic bi-modal datasets become available in the future.

\section{Threats to Validity \& Limitations}\label{sec:threats}
\noindent\textbf{Construct validity:} Our COOD/COOD+ framework uses data-driven techniques to synthesize OOD samples, which may not fully reflect real-world SE scenarios. While we include diverse OOD scenarios, a pilot study with developers is necessary. Additionally, the reliability of our OOD benchmark depends heavily on the quality of the OOD datasets used. \textbf{Internal validity:} Hyperparameter tuning impacts ML performance. For model fine-tuning, we kept the GraphCodeBERT architecture unchanged due to feasibility reasons, but conducted ablation studies with various model components, encoder backbones, and key hyperparameters. \textbf{External validity:} We conduct OOD detection experiments on two large-scale code search datasets. Although our focus on Python and Java limits generalizability, experiments on these two languages partially demonstrate that our approach is PL-agnostic.

\section{Conclusion}\label{sec:conclusion} 
We proposed two multi-modal OOD detection aproaches for code-related pre-trained ML models; namely unsupervised COOD and weakly-supervised COOD+. The COOD merely leveraged unsupervised contrastive learning to identify OOD samples. As an extension of COOD, COOD+ combined contrastive learning and a binary classifier for OOD detection using a small number of labelled OOD samples. To reap the benefits of these two modules, we also devised a novel scoring metric to fuse their prediction results. The evaluation results demonstrated that the integration of the rejection network and contrastive learning can achieve superior performance in detecting all four OOD scenarios for multi-modal NL-PL data. Additionally, our models can be applied to the downstream SE task, achieving comparable performance to existing code-related models. 


\section{Acknowledgement}\label{sec:acknowledgement}
This research has been supported in part by the NSF CCF-2311469, CNS-2132281, CCF-2007246, CCF-1955853 and CNS-2346357. We also gratefully acknowledge support from Cisco Systems. The opinions, findings, and conclusions expressed in this work are solely those of the authors and do not necessarily reflect the views of the sponsors.
\bibliographystyle{IEEEtran}
\bibliography{IEEEexample, main_no_url}

\end{document}